\documentclass[letterpaper,twocolumn,english,superscriptaddress,prl,aps,showpacs]{revtex4}
\usepackage[T1]{fontenc}
\usepackage[latin1]{inputenc}
\usepackage{amsmath}
\usepackage{graphicx}
\usepackage{amssymb}
\usepackage{esint}

\makeatletter
\@ifundefined{textcolor}{}
{%
 \definecolor{BLACK}{gray}{0}
 \definecolor{WHITE}{gray}{1}
 \definecolor{RED}{rgb}{1,0,0}
 \definecolor{GREEN}{rgb}{0,1,0}
 \definecolor{BLUE}{rgb}{0,0,1}
 \definecolor{CYAN}{cmyk}{1,0,0,0}
 \definecolor{MAGENTA}{cmyk}{0,1,0,0}
 \definecolor{YELLOW}{cmyk}{0,0,1,0}
 }

\@ifundefined{definecolor}{\@ifundefined{definecolor}
 {\@ifundefined{definecolor}
 {\@ifundefined{definecolor}
 {\usepackage{color}}{}
}{}
}{}
}{}\makeatother

\makeatother

\usepackage{babel}\DeclareMathAlphabet{\mathpzc}{OT1}{pzc}{m}{it}

\makeatother

\usepackage{babel}

\makeatother

\usepackage{babel}

\begin{document}

\author{Alexey A. Kovalev}

\affiliation{Department of Physics and Astronomy, University of California, Los
Angeles, California 90095, USA}

\author{Lorien X. Hayden}

\affiliation{Department of Physics, University of Missouri, Columbia, MO 65211, USA}

\author{Gerrit E. W. Bauer}

\affiliation{Institute for Materials Research, Tohoku University, Sendai 980-8577, Japan}
\affiliation{Kavli Institute of NanoScience, Delft University of Technology, Lorentzweg
1, 2628 CJ Delft, The Netherlands}

\author{Yaroslav Tserkovnyak}

\affiliation{Department of Physics and Astronomy, University of California, Los
Angeles, California 90095, USA}

\title{Macrospin Tunneling and Magnetopolaritons with Nanomechanical Interference}
\begin{abstract}
We theoretically address the quantum dynamics of a nanomechanical
resonator coupled to the macrospin of a magnetic nanoparticle by both
instanton and perturbative approaches. We demonstrate suppression
of the tunneling between opposite magnetizations and destruction of magnetopolaritons (coherent magneto-mechanical oscillations)  by nanomechanical
interference.
The predictions can be verified experimentally by a molecular magnet
attached to a nanomechanical bridge. 
\end{abstract}

\date{\today}

\pacs{75.80.+q, 75.50.Xx, 85.65.+h, 75.45.+j}

\maketitle
The first direct observation of quantum behavior of a macroscopic
mechanical resonator constituting a nano-electro-mechanical system
(NEMS) has been reported recently \cite{O'Connell:apr2010}. This
opens a wide range of new possibilities for testing quantum-mechanical
principles on macroscopic objects and has the potential to impact
sensor technology. NEMS's have also been suggested to operate as qubits
and memory elements for quantum-information processing \cite{Cleland:AUG2004}.
Proposals and realizations
of two-level systems (e.g., superconducting qubits) coupled to mechanical modes \cite{Cleland:AUG2004,Hauss:2008} allow quantum
measurements on the mechanical resonator. Here, we study quantum effects
in a NEMS coupled to a ferromagnetic
nanoparticle such as a single-molecule magnet (SMM).

The dynamics of a magnetic order parameter and a mechanical resonator
are coupled by conservation of angular momentum \cite{Kovalev:AUG2003}.
The magnetization dynamics of a ferromagnetic particle \cite{Kovalev:AUG2003}
as well as macrospin tunneling oscillations in an SMM \cite{Jaafar:Jun2009}
should in principle induce magnetomechanical motion. However, the semiclassical
treatment fails when the coupling becomes stronger and
quantum mechanical effects such as freezing of spin tunneling
\cite{Chudnovsky:Jun2010} manifest themselves, as discussed below.

\begin{figure}
\includegraphics[clip,width=1\linewidth]{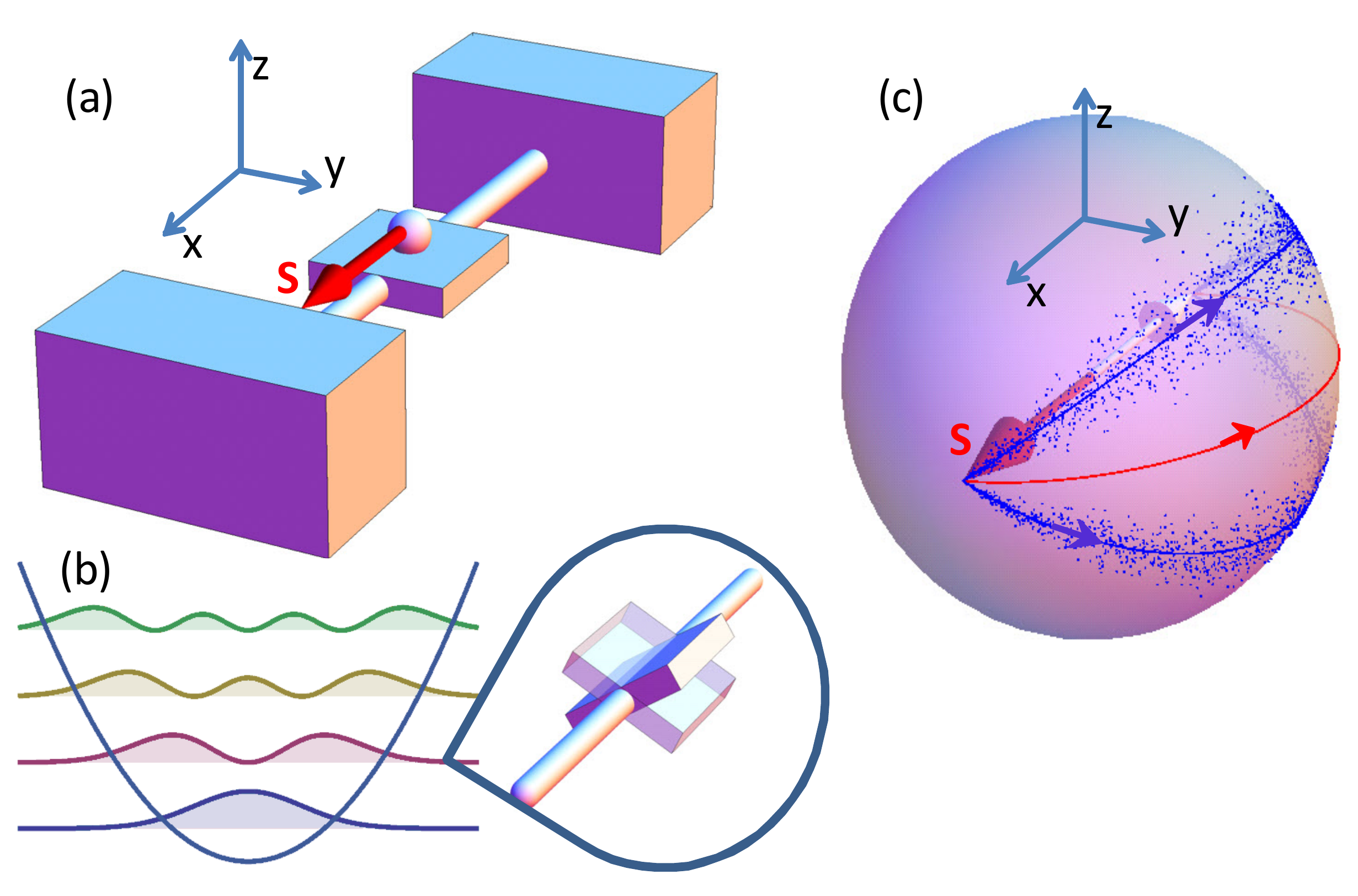} \caption{(a) A torsional resonator consisting of a beam (e.g., nanotube) and
a load (e.g., a magnetic nanoparticle or an SMM attached to a paddle).
(b) Probability of finding the resonator at a given angle for the
four lowest energy levels. (c) The first excited state of the resonator
is effectively a superposition of two states with positive and negative
torsion angles. The spin-reversal tunnel path on a unit sphere with
a rigid resonator (middle geodesic) splits into two equivalent ones
(fuzzy paths represent uncertainty in the tilt of the mechanical resonator),
allowing interference.}

\label{fig1} 
\end{figure}

In this Letter, we consider a torsional nanomechanical resonator {[}see
Fig.~\ref{fig1}(a){]} consisting of a load (e.g., a magnetic nanoparticle or an SMM
attached to a paddle) and a mechanical link to the base (e.g., a nanotube
or a chemical bond). The projection of the wave function of the macrospin on the two lowest energy levels is equivalent
to a harmonic oscillator coupled to a two-level system \cite{Jaynes:jan.1963}.
The interference effects discussed here can be understood by considering
a mechanical resonator in the $n$th excited state {[}see Fig.~\ref{fig1}(b){]},
which has $n+1$ probability maxima at different torsion angles. For
example, the first excited state can be thought of as a superposition
of two wave functions peaked at different torsion angles \cite{Note}. 

For the first excited state, the instanton path on a unit sphere {[}the
middle geodesic in Fig.~\ref{fig1}(c){]} is split into two equivalent
ones {[}geodesics with a scatter in Fig.~\ref{fig1}(c){]} due to
magnetic anisotropies defining the tunneling trajectory (e.g. an easy
$xy-$plane anisotropy). The area between the equivalent paths in
Fig.~\ref{fig1}(c) multiplied by the spin $S$ is equal to the difference
in the phases accumulated by the two paths, which leads to a complete
suppression of tunneling for a phase difference of $\pi$. Interference
effects are therefore observable when $S\sqrt{\hbar/I_{x}\omega_{r}}\sim\pi$,
where $I_{x}$ is the moment of inertia of the load and $\omega_{r}$
is the natural frequency of the resonator (the feasibility of this
regime will be discussed below). We find that interference is most
significant in systems, in which there is a certain ratio between
the spin and mechanical angular momenta which is analogous to the
selection (parity) rules in large-angle macrospin-tunneling \cite{Wernsdorfer:MAY2000}.
We predict that the effect is rather robust and can be observed at
experimentally achievable temperatures in state-of-the-art structures.
Furthermore, tunneling can be suppressed by raising temperature (thus
repopulating the lowest states but without increasing decoherence). Below,
we derive rigorously that quantum-mechanical oscillations of the resonator
indeed lead to a suppression of macrospin tunneling and destruction
of magnetomechanical modes. Using the instanton approach, we find
that the coupling of a magnetic particle with an easy-plane anisotropy
to a mechanical resonator can only lower the tunneling rate, thus
stabilizing the spin.

Consider a magnetic nanoparticle that behaves as a rigid, spin-$S$ object and is characterized by
the magnetic anisotropy energy $\hat{H}_{A}=E(\hat{S}_{x},\hat{S}_{y},\hat{S}_{z})$,
where $\hat{S}_{x,(y,z)}$ are the spin-projection operators (in units
of $\hbar$). Here, $E(S_{x},S_{y},S_{z})$ is the classical magnetic
energy corresponding to an easy $x$ axis and a transverse perturbation (parity symmetric about the $y-z$ plane)
that couples the magnet to the torsional motion (e.g. an easy $xy-$plane
anisotropy). Our complete Hamiltonian is \begin{equation}
\hat{H}=\hat{H}_{r}+e^{-i\hat{S}_{x}\hat{\varphi}}\hat{H}_{A}e^{i\hat{S}_{x}\hat{\varphi}}+\gamma\hbar\hat{S}_{x}B\,,\label{Hamiltonian}\end{equation}
 where the spin is coupled to a single mechanical mode with frequency
$\omega_{r}$ and Hamiltonian $\hat{H}_{r}=\hbar\omega_{r}(\hat{a}^{\dagger}\hat{a}+1/2)$
in terms of creation/annihilation operators $\hat{a}^{\dagger}$/$\hat{a}$.
The second term in Eq.~(\ref{Hamiltonian}) describes magnetic anisotropy,
taking into account its orientation with respect to the lattice in
terms of the torsion angle $\hat{\varphi}=(\alpha\hat{a}^{\dagger}+\alpha^{*}\hat{a})/2S$
($\alpha=S\sqrt{2\hbar/I_{x}\omega_{r}}$ for the system in Fig. \ref{fig1};
however, the analytical solution presented below will also hold for
an arbitrary complex $\alpha$). The third term is due to the external
magnetic field $B$ along the $x$ axis ($\gamma$ is minus the gyromagnetic
ratio). We now transform the Hamiltonian (\ref{Hamiltonian}) by a
unitary transformation $e^{-i\hat{S}_{x}\hat{\varphi}}$ to the rest
frame:\begin{equation}
\hat{H}_{R}=\hbar\omega_{r}(\hat{\tilde{a}}^{\dagger}\hat{\tilde{a}}+1/2)+\hat{H}_{A}+\gamma\hbar\hat{S}_{x}B\,.\label{Ham1}\end{equation}
 Here, $\hat{\tilde{a}}=\hat{a}-i\hat{S}_{x}\alpha/2S$. We describe
tunneling between low-lying states of the macrospin by path integrals
in which coherent states are constructed using the Heisenberg-Weyl
(the resonator) and SU(2) (the spin) groups from a state $\left|\uparrow,n\right\rangle $
by a standard procedure \cite{Perelomov:1986}: \begin{equation}
\left|\boldsymbol{\Omega},z\right\rangle =e^{z\hat{a}^{\dagger}-z^{*}\hat{a}}e^{-i\hat{S}_{z}\phi}e^{-i\hat{S}_{y}\theta}\left|\uparrow,n\right\rangle \,.\label{CoherentState}\end{equation}
 Index $n$ here stands for a Fock state with $n$ phonons in the
mechanical mode; $\uparrow$ refers to a macrospin state pointing
to the north pole; $\theta$ and $\phi$ are the Euler angles defining
direction $\boldsymbol{\Omega}=(\theta,\phi)$ of the macrospin; and
the complex-valued $z=z_{r}+iz_{i}$ parametrizes a generalized coherent
state of the harmonic oscillator.

In the large-S limit, the transition amplitude between two approximate eigenstates at $\theta=\pi/2,\,\phi=0,\, z=i\alpha/2$
and $\theta=\pi/2,\,\phi=\pi,\, z=-i\alpha/2$ can be expressed
through coherent-state path integrals \cite{Chudnovsky:Feb1988} as \begin{equation}
\left\langle -\mathbf{x},-i\alpha/2\left|e^{-i\int dt\hat{H}}\right|\mathbf{x},i\alpha/2\right\rangle ={\displaystyle \int\mathcal{D}\Omega e^{i(\mathcal{S}_{k}+\mathcal{S}_{E})/\hbar}}\,,\label{Tamplitude}\end{equation}
where $\mathcal{D}\Omega\sim\prod_{t}d\phi_{t}d(\cos\theta_{t})dz_{t}dz_{t}^{*}$,
$\mathcal{S}_{k}=\hbar\int dt\left[(\dot{z}_{r}z_{i}-\dot{z}_{i}z_{r})-S\dot{\phi}(1-\cos\theta)\right]$
is the kinetic and $\mathcal{S}_{E}=-\int dt\tilde{E}$, where $\tilde{E}=\hbar\omega_{r}\left[n+1/2+(z-is_{x}\alpha/2)(z^{*}+is_{x}\alpha/2)\right]+E(\theta,\phi)$,
the potential energy contribution to the action in the absence of
a magnetic field ($s_{x}=\sin\theta\cos\phi$). The term $\mathcal{S}_{k}$
describes interference effects that can be exposed by treating the
transition amplitude in Eq.~(\ref{Tamplitude}) by a saddle-point
approximation. Each saddle-point path acquires a phase $SA+2A'$,
where $A$ is the solid angle spanned by the spin paths connected
to the north pole by geodesics and $A'$ is the area enclosed by the
torsional trajectories connected to the origin in the complex plane.
These phases are known to cause interference effects in the tunneling
of spins \cite{Loss:Nov1992}.

\begin{figure}
\includegraphics[clip,width=1\linewidth]{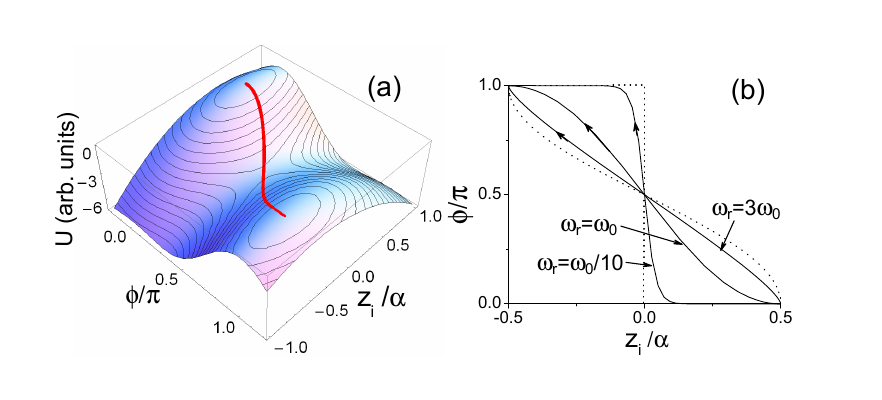} \caption{(a) A particle falls in a camel-back shaped potential following the
quasiclassical path (bold curve). (b) Results of a numerical integration
of Eqs.~(\ref{EqMotion}) for different frequencies of the mechanical
resonator.}

\label{fig3} 
\end{figure}

We first calculate the tunneling rate in Eq. (\ref{Tamplitude}) by
a quasiclassical treatment in imaginary time. The quasiclassical equations
of motion in real time minimize the action in Eq.~(\ref{Tamplitude}) \cite{Note1}:
\begin{align}
\dot{z}_{i} & =-\omega_{r}z_{r}\,,\,\,\,\dot{z}_{r}=\omega_{r}z_{i}-(\alpha\omega_{r}/2)\cos\phi\sin\theta\,,\nonumber \\
S\dot{\theta}\sin\theta & =\partial\tilde{E}/\partial\phi\,,\,\,\, S\dot{\phi}\sin\theta=-\partial\tilde{E}/\partial\theta\,.\label{EqMotion}\end{align}
In order to find the instanton path in the presence of the coupling
to the mechanical mode, we integrate these equations numerically in
imaginary time ($t=-i\tau$). The splitting of the degenerate modes
by the tunnel interaction can be expressed as: \begin{equation}
\Delta_{n}\propto\mathcal{C}e^{\mathrm{Re}\left(\mathcal{S}_{0}^{cl}\right)/\hbar}\,,\label{QCsplitting}\end{equation}
where $\mathcal{S}_{0}^{cl}=i\mathcal{S}_{k}+\mathcal{S}_{E}$ is the Wick rotated
instanton action of a quasiclassical trajectory found by solving Eqs.
(\ref{EqMotion}) for $\theta$, $\phi$, $z_{i}$, and $z_{r}$.
$\mathcal{C}$ describes the interference of different spin trajectories
under tunneling selection rules (e.g. for an easy $xy-$plane anisotropy
$\mathcal{C}=\cos(\pi S)$) \cite{Loss:Nov1992}.
For an easy-axis/easy-plane anisotropy described by $E=K_{1}\cos^{2}\theta+K_{2}\sin^{2}\theta\sin^{2}\phi$
($K_{1}>K_{2}>0$), we find that all paths are contained between two
extremal paths denoted in Fig.~\ref{fig3}(b) by dotted lines. The
tunnel splitting changes from $\exp(-\alpha^{2}/2)\Delta_{0}$ {[}rectangular-shaped
dotted path in Fig.~\ref{fig3}(b){]} to $\Delta_{0}$ {[}cosine-shaped
dotted path in Fig.~\ref{fig3}(b){]} as we go from the limit $\omega_{r}\ll\omega_{0}\sim t_{I}^{-1}$
(which is the focus of this Letter) to the limit $\omega_{r}\gg\omega_{0}\sim t_{I}^{-1}$
where $\Delta_{0}$ is the tunnel splitting without coupling to the
resonator. $\omega_{0}=S\sqrt{K_{1}K_{2}}$ and $t_{I}$ is the uncoupled
(magnetic) instanton tunneling time. Such behavior can be understood
by eliminating $\theta$ and $z_{r}$ from Eqs.~(\ref{EqMotion})
and the imaginary-time Lagrangian: \begin{equation}
L=M_{1}(\phi)\dot{\phi}^{2}/2+M_{2}\dot{z}_{i}^{2}/2-U(\phi,z)\,.\end{equation}
Here $M_{1}(\phi)=\omega_{0}^{-1}(1/\sqrt{\lambda}-\sqrt{\lambda}\sin^{2}\phi+\alpha\cos\phi\omega_{r}/2\omega_{0})$,
$M_{2}=2/\omega_{r}$, $U(\phi,z)=-\sqrt{\lambda}\omega_{0}\sin^{2}\phi-\omega_{r}(z_{i}-\cos\phi\alpha/2)^{2}$,
$\lambda=K_{2}/K_{1}$, and $\theta\approx\pi/2$ has been used. By
inspecting the motion of a particle with anisotropic mass in the potential
$U(\phi,z)$ {[}Fig.~\ref{fig3}(a){]}, we see that the quasiclassical
path connecting two potential-energy maxima leads to a smaller tunnel
splitting because it experiences a higher tunneling barrier compared
to an uncoupled instanton (this is also expected for other magnetic anisotropies).

\begin{figure}
\includegraphics[clip,width=1\linewidth]{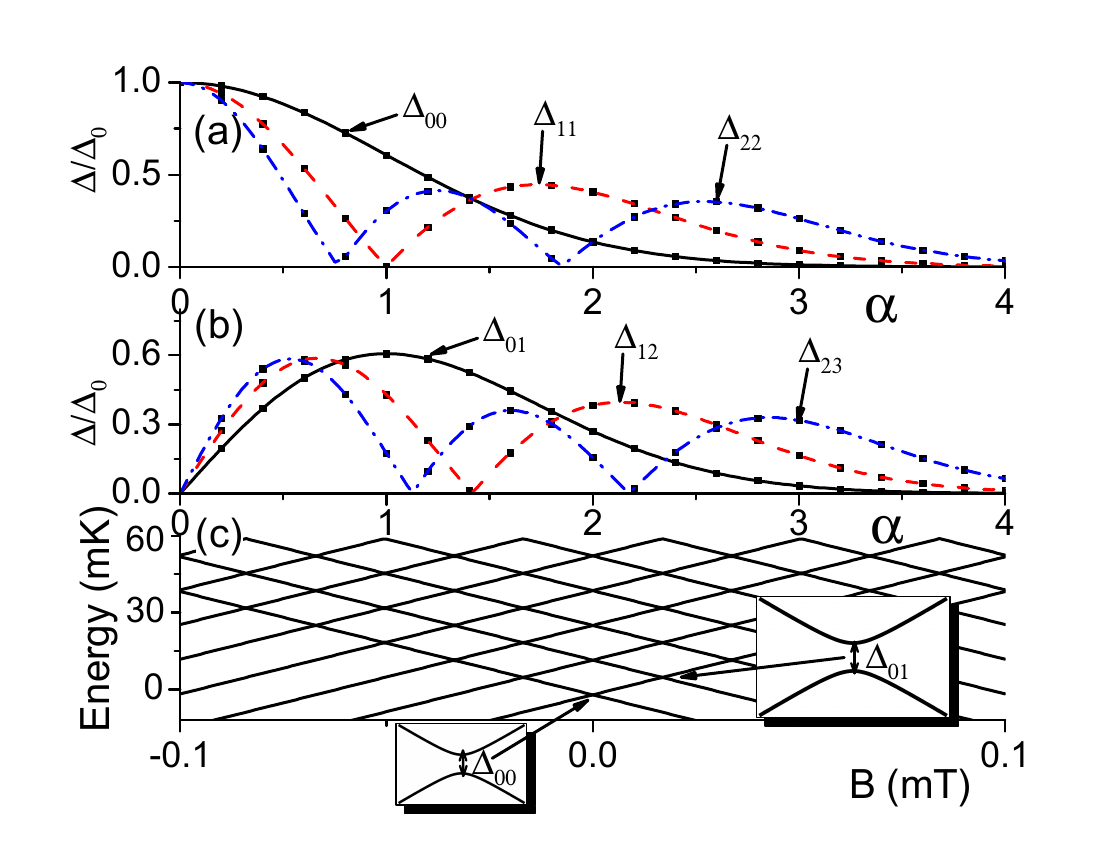} \caption{(a) Tunnel splittings as a function of the macrospin-resonator coupling
$\alpha$ for the first three excited states of the resonator. The
curves show analytical results, while the squares are based on the
numerical diagonalization of the Hamiltonian corresponding to an Fe$_{8}$
SMM. (b) Analogous plot for tunnel splittings of the magnetopolariton
modes corresponding to the Fock states differing by one phonon. (c)
Lowest energy levels of the Fe$_{8}$ SMM coupled to a mechanical
resonator obtained by numerical diagonalization. $\Delta_{0}/\hbar\omega_{r}=3\times10^{-6}$
(the energy is offset by $-28.239$ K).}

\label{fig2} 
\end{figure}

We show in the following that the quasiclassical approach only works
well for $n=0$ but does not capture nanomechanical interference effects.
The latter can be obtained by calculating the quantum fluctuations
and by retaining the second order terms (depending on the Fock number
$n$) in the kinetic-energy of the Lagrangian. In the remaining
part of this Letter, we consider the limit $t_{I}\ll\omega_{r}^{-1}$
in which the kinetic-energy contribution to the action suppresses
the resonator dynamics. The resonator
contribution can then be calculated by
taking the matrix element between the initial and final states of
the mechanical subsystem: \begin{equation}
\left\langle -\mathbf{x},-i\alpha/2\left|e^{-i\int dt\hat{H}}\right|\mathbf{x},i\alpha/2\right\rangle =\kappa_{nn}\int\mathcal{D}\Omega e^{i\mathcal{S}_{I}/\hbar}\,.\label{SingleInstanton}\end{equation}
Here, $\kappa_{nm}=\left\langle n\left|e^{-i2S\hat{\varphi}}\right|m\right\rangle $
is the Fock states' matrix element of the displacement operator and
$e^{i\mathcal{S}_{I}/\hbar}$ corresponds to the bare instanton contribution
to the path integral (without the coupling to the resonator). The
extra factor in the instanton contribution reflects the phases accumulated
by multiple paths of the macrospin tunneling within the laboratory
frame in Eq.~(\ref{Hamiltonian}). These paths can destructively
interfere, suppressing tunneling at specific values of $\alpha$ {[}see
Fig.~\ref{fig2}(a){]}, as becomes clear from the following analytical
expression \cite{Cahill:1969}:\begin{equation}
\kappa_{nm}=e^{-|\alpha|^{2}/2}(-i\alpha)^{n-m}\sqrt{\dfrac{m!}{n!}}L_{m}^{(n-m)}(|\alpha|^{2})\,,\label{Laguerre}\end{equation}
where $n\geq m$ ($n<m$ can be obtained by complex conjugation) and
$L_{m}^{(n-m)}$ is a generalized Laguerre polynomial. We can then
anticipate that the tunnel splitting should be renormalized by the
nanoresonator according to: \begin{equation}
\Delta_{n}=\left|\kappa_{nn}\right|\Delta_{0}\leq\Delta_{0}\,,\label{Splitting}\end{equation}
where $\Delta_{0}$ is the tunnel splitting for a bare macrospin.
A more rigorous derivation also applicable to the formation of magnetopolaritons
($m\neq n$) is discussed below.

To tackle the resonant coupling between Fock states with arbitrary numbers of phonons, we project
Hamiltonian (\ref{Ham1}) onto the basis formed by the two lowest
energy states of the Hamiltonian $\hat{H}_{A}$ which is possible when the transverse perturbations are small (i.e. when $\Delta_0\ll\omega_0$ -- the distance to the third energy state). We can represent
the ground and first excited states split by $\Delta_{0}$ as \cite{Chudnovsky:Jun2010}
\begin{equation}
\Psi_{\pm}=\dfrac{1}{\sqrt{2}}\left(|{\psi_{\mathbf{x}}}\rangle\pm|{\psi_{-\mathbf{x}}}\rangle\right)\,,\end{equation}
 where states $|{\psi_{\pm\mathbf{x}}}\rangle$ represent perturbations
of $|{\pm\mathbf{x}}\rangle$. The leading-order projection procedure, \begin{equation}
\hat{H}_{p}=\sum_{\mu,\nu=\pm\mathbf{x}}\left\langle \psi_{\mu}\right|\hat{H}\left|\psi_{\nu}\right\rangle \left|\psi_{\mu}\right\rangle \left\langle \psi_{\nu}\right|\,,\label{Projection}\end{equation}
 leads to the Hamiltonian \cite{Chudnovsky:Jun2010}:\begin{equation}
\hat{H}_{p}=\left[\begin{array}{cc}
\hat{H}_{r}+\gamma\hbar SB & -e^{-i2S\hat{\varphi}}\Delta_{0}/2\\
-e^{i2S\hat{\varphi}}\Delta_{0}/2 & \hat{H}_{r}-\gamma\hbar SB\end{array}\right]\,,\label{SAhamiltonian}\end{equation}
 where we use $\left\langle \psi_{m}\right|\hat{S}_{x}\left|\psi_{n}\right\rangle \cong\pm S\delta_{mn}$
for states $|{\psi_{\pm\mathbf{x}}}\rangle$. In Eq. (\ref{SAhamiltonian})
we treat $\triangle_{0}$ as a small perturbation. Therefore, in the
vicinity of the resonant magnetic field corresponding to the crossing
of Fock states $m$ and $n$, we can further project Eq.~(\ref{SAhamiltonian})
onto the states $m$ and $n$: \begin{equation}
\hat{H}_{nm}=\left[\begin{array}{cc}
E_{n} & -\kappa_{nm}\Delta_{0}/2\\
-\kappa_{nm}^{*}\Delta_{0}/2 & E_{m}\end{array}\right]\,,\end{equation}
 arriving at eigenenergies $E_{nm}=(E_{n}+E_{m})/2\pm\sqrt{(E_{n}-E_{m})^{2}+\Delta_{nm}^{2}}/2$,
where the energy of the Fock state $n$ $(m)$ is $E_{n}=\hbar\omega_{r}(n+1/2)+\gamma\hbar SB$
{[}$E_{m}=\hbar\omega_{r}(m+1/2)-\gamma\hbar SB${]} and $\Delta_{mn}=|\kappa_{mn}|\Delta_{0}$.
From the eigenenergies, we can immediately see that the magnetopolariton
and tunnel splittings are given by $\Delta_{mn}$, which is the main
result of this Letter.

Finally, we confirm our analytical results by an exact diagonalization
of Hamiltonian (\ref{Hamiltonian}) in the basis that includes $30$
Fock states for spin $S=10$, which corresponds to a $630\times630$
matrix. A typical spin Hamiltonian describing an SMM reads \begin{equation}
\hat{H}_{A}=-D\hat{S}_{x}^{2}+E(\hat{S}_{z}^{2}-\hat{S}_{y}^{2})+C(\hat{S}_{+}^{4}+\hat{S}_{-}^{4})\,,\end{equation}
 where $S_{\pm}=S_{z}\pm iS_{y}$. The anisotropy constants were taken
to correspond to an Fe$_{8}$ single-molecule magnet with $D=0.292$~K,
$E=0.046$~K, and $C=-2.9\times10^{-5}$~K \cite{Wernsdorfer:MAY2000}.
Such parameters result in the tunnel splitting $\Delta_{0}/k_{B}=4.5\times10^{-8}$~K
in the absence of coupling to the resonator. The resonator frequency
is chosen as $\hbar\omega_{r}/k_{B}=3\times10^{5}\Delta_{0}/k_{B}\sim15$
mK where $k_{B}$ is the Boltzmann constant. In Fig.~\ref{fig2}(c), we plot the calculated lowest eigenenergies.
The energy (anti-) crossings at zero field corresponding to the tunnel splitting
are shown in Fig.~\ref{fig2}(a) as a function of the coupling parameter
$\alpha$. We observe a perfect agreement between the results of Eq.~(\ref{Splitting})
(lines) and the results of the numerical diagonalization (squares).
At specific values of $\alpha$, we observe destructive interference
that completely quenches the macrospin tunneling. The tunnel splittings
in Fig.~\ref{fig2}(c) at nonzero magnetic fields correspond to magnetopolariton
formation and are only possible for a finite coupling to the resonator.
The magnetopolariton splittings reveal interference effects as a function
of this coupling (see Fig.~\ref{fig2}(b) for analytical (lines)
as well as numerical (squares) results).

The tunnel splittings can be measured by the Landau-Zener method employed
in Ref.~\cite{Wernsdorfer:MAY2000}. The mechanical resonator has
to be cooled to temperatures $T\ll\hbar\omega_{r}/k_{B}$ (thus higher frequency resonators are preferable); e.g. by quantum-optical \cite{Rocheleau:jan2010}
or conventional cryogenic techniques (in Ref. \cite{O'Connell:apr2010} $\hbar\omega_{r}/k_{B}\sim0.1$ K). The
critical parameter for the observation of magnetopolariton modes and
interference effects is the macrospin-resonator coupling $\alpha$.
We estimate its value for a device that contains an SMM strongly absorbed
to a paddle of the size $20\times20\times10\:\mbox{nm}^{3}$, with
a single-wall carbon nanotube serving as a mechanical link {[}see
Fig.~\ref{fig1}(a){]}. For a torsional spring constant $K=10^{-18}$~N$\cdot$m
\cite{Meyer:SEP2005}, $S=37$ (for a Mn$_{17}$ SMM \cite{Moushi:SEP2009})
and a moment of inertia $I_{x}=10^{-36}$~kg$\cdot$m$^{2}$, we
obtain $\omega_{r}=\sqrt{K/I_{x}}\sim1$~GHz and $\alpha\sim0.02$.
This coupling is large enough for the observation of magnetopolariton
modes but too small for observing interference effects for which $\alpha$
has to be comparable to $0.5$ {[}see Fig.~\ref{fig2}(a){]}. Sufficiently
large values of the coupling can be achieved in a Mn$_{12}$ SMM bridged
between leads \cite{Barraza-Lopez:Jun2009}. For parameters similar
to the ones used in Ref.~\cite{Jaafar:Jun2009}, $I_{x}\sim10^{-41}$~kg$\cdot$m$^{2}$,
$\omega_{r}\sim1$~GHz and $S=10$, we arrive at $\alpha\sim1.5$.
The spin-resonator coupling can be increased by lowering the torsional
stiffness and the moment of inertia or by increasing the spin $S$.
The energy levels in Fig.~\ref{fig2}(c) can also be used for quantum
manipulations and single-phonon control of a mechanical resonator.
This, however, requires larger tunnel splittings in order to overcome
decoherence. The tunnel splitting can be easily increased
by applying the magnetic field normal to the anisotropy axis \cite{Barco:SEP1999}.

To conclude, we found a quantum-mechanical solution for the coupled
motion of a macrospin and a mechanical resonator.
We study tunnel splittings and avoided level crossings corresponding
to formation of magnetopolaritons, both of which should be detectable by quantum
optical techniques or by studying Landau-Zener transitions \cite{Wernsdorfer:MAY2000}.
In the strong spin-resonator coupling regime, we predict suppression
of the tunneling of magnetization and destruction of magnetopolaritons
by interference of the spin tunneling paths resulting from the quantum
state of the resonator. We predict that the magnetism in SMMs can
be significantly stabilized against quantum fluctuations by sticking
them to mechanical resonators with large quantum fluctuations. Results
presented here are relevant for possible realizations of quantum control
of magnetization at a single-phonon level.

We thank DARPA, Alfred P. Sloan Foundation,
NSF-DMR-0840965 (Y. T.) and NSF-PHY-0850501
(L. H.), and the Dutch FOM foundation (G. B.).

\bibliographystyle{apsrev}
\bibliography{mnems}

\begin{thebibliography}{23}
\expandafter\ifx\csname natexlab\endcsname\relax\def\natexlab#1{#1}\fi
\expandafter\ifx\csname bibnamefont\endcsname\relax
  \def\bibnamefont#1{#1}\fi
\expandafter\ifx\csname bibfnamefont\endcsname\relax
  \def\bibfnamefont#1{#1}\fi
\expandafter\ifx\csname citenamefont\endcsname\relax
  \def\citenamefont#1{#1}\fi
\expandafter\ifx\csname url\endcsname\relax
  \def\url#1{\texttt{#1}}\fi
\expandafter\ifx\csname urlprefix\endcsname\relax\def\urlprefix{URL }\fi
\providecommand{\bibinfo}[2]{#2}
\providecommand{\eprint}[2][]{\url{#2}}

\bibitem[{\citenamefont{O'Connell et~al.}(2010)\citenamefont{O'Connell,
  Hofheinz, Ansmann, Bialczak, Lenander, Lucero, Neeley, Sank, Wang, Weides
  et~al.}}]{O'Connell:apr2010}
\bibinfo{author}{\bibfnamefont{O.~D.} \bibnamefont{O'Connell}},
  \bibinfo{author}{\bibfnamefont{M.}~\bibnamefont{Hofheinz}},
  \bibinfo{author}{\bibfnamefont{M.}~\bibnamefont{Ansmann}},
  \bibinfo{author}{\bibfnamefont{R.~C.} \bibnamefont{Bialczak}},
  \bibinfo{author}{\bibfnamefont{M.}~\bibnamefont{Lenander}},
  \bibinfo{author}{\bibfnamefont{E.}~\bibnamefont{Lucero}},
  \bibinfo{author}{\bibfnamefont{M.}~\bibnamefont{Neeley}},
  \bibinfo{author}{\bibfnamefont{D.}~\bibnamefont{Sank}},
  \bibinfo{author}{\bibfnamefont{H.}~\bibnamefont{Wang}},
  \bibinfo{author}{\bibfnamefont{M.}~\bibnamefont{Weides}},
  \bibnamefont{et~al.}, \bibinfo{journal}{Nature}
  \textbf{\bibinfo{volume}{464}}, \bibinfo{pages}{697} (\bibinfo{year}{2010}).

\bibitem[{\citenamefont{Cleland and Geller}(2004)}]{Cleland:AUG2004}
\bibinfo{author}{\bibfnamefont{A.~N.} \bibnamefont{Cleland}} \bibnamefont{and}
  \bibinfo{author}{\bibfnamefont{M.~R.} \bibnamefont{Geller}},
  \bibinfo{journal}{Phys. Rev. Lett.} \textbf{\bibinfo{volume}{93}},
  \bibinfo{pages}{070501} (\bibinfo{year}{2004});
\bibinfo{author}{\bibfnamefont{M.~R.} \bibnamefont{Geller}} \bibnamefont{and}
  \bibinfo{author}{\bibfnamefont{A.~N.} \bibnamefont{Cleland}},
  \bibinfo{journal}{Phys. Rev. A} \textbf{\bibinfo{volume}{71}},
  \bibinfo{pages}{032311} (\bibinfo{year}{2005});
\bibinfo{author}{\bibfnamefont{S.}~\bibnamefont{Savel'ev}},
  \bibinfo{author}{\bibfnamefont{X.}~\bibnamefont{Hu}}, \bibnamefont{and}
  \bibinfo{author}{\bibfnamefont{F.}~\bibnamefont{Nori}}, \bibinfo{journal}{New
  J. Phys.} \textbf{\bibinfo{volume}{8}}, \bibinfo{pages}{105}
  (\bibinfo{year}{2006}).

\bibitem[{\citenamefont{Hauss et~al.}(2008)\citenamefont{Hauss, Fedorov,
  Andr\'{e}, Brosco, Hutter, Kothari, Yeshwanth, Shnirman, and
  Sch\"{o}n}}]{Hauss:2008}
\bibinfo{author}{\bibfnamefont{J.}~\bibnamefont{Hauss}},
  \bibinfo{author}{\bibfnamefont{A.}~\bibnamefont{Fedorov}},
  \bibinfo{author}{\bibfnamefont{S.}~\bibnamefont{Andr\'{e}}},
  \bibinfo{author}{\bibfnamefont{V.}~\bibnamefont{Brosco}},
  \bibinfo{author}{\bibfnamefont{C.}~\bibnamefont{Hutter}},
  \bibinfo{author}{\bibfnamefont{R.}~\bibnamefont{Kothari}},
  \bibinfo{author}{\bibfnamefont{S.}~\bibnamefont{Yeshwanth}},
  \bibinfo{author}{\bibfnamefont{A.}~\bibnamefont{Shnirman}}, \bibnamefont{and}
  \bibinfo{author}{\bibfnamefont{G.}~\bibnamefont{Sch\"{o}n}},
  \bibinfo{journal}{New J. Phys.} \textbf{\bibinfo{volume}{10}},
  \bibinfo{pages}{095018} (\bibinfo{year}{2008});
\bibinfo{author}{\bibfnamefont{A.~D.} \bibnamefont{Armour}} \bibnamefont{and}
  \bibinfo{author}{\bibfnamefont{M.~P.} \bibnamefont{Blencowe}},
  \bibinfo{journal}{New J. Phys.} \textbf{\bibinfo{volume}{10}},
  \bibinfo{pages}{095004} (\bibinfo{year}{2008});
\bibinfo{author}{\bibfnamefont{M.~D.} \bibnamefont{LaHaye}},
  \bibinfo{author}{\bibfnamefont{J.}~\bibnamefont{Suh}},
  \bibinfo{author}{\bibfnamefont{P.~M.} \bibnamefont{Echternach}},
  \bibinfo{author}{\bibfnamefont{K.~C.} \bibnamefont{Schwab}},
  \bibnamefont{and} \bibinfo{author}{\bibfnamefont{M.~L.}
  \bibnamefont{Roukes}}, \bibinfo{journal}{Nature}
  \textbf{\bibinfo{volume}{459}}, \bibinfo{pages}{960} (\bibinfo{year}{2009}).

\bibitem[{\citenamefont{Kovalev et~al.}(2003)\citenamefont{Kovalev, Bauer, and
  Brataas}}]{Kovalev:AUG2003}
\bibinfo{author}{\bibfnamefont{A.~A.} \bibnamefont{Kovalev}},
  \bibinfo{author}{\bibfnamefont{G.~E.~W.} \bibnamefont{Bauer}},
  \bibnamefont{and} \bibinfo{author}{\bibfnamefont{A.}~\bibnamefont{Brataas}},
  \bibinfo{journal}{Appl. Phys. Lett.} \textbf{\bibinfo{volume}{83}},
  \bibinfo{pages}{1584} (\bibinfo{year}{2003});
\bibinfo{author}{\bibfnamefont{A.~A.} \bibnamefont{Kovalev}},
  \bibinfo{author}{\bibfnamefont{G.~E.~W.} \bibnamefont{Bauer}},
  \bibnamefont{and} \bibinfo{author}{\bibfnamefont{A.}~\bibnamefont{Brataas}},
  \bibinfo{journal}{Phys. Rev. Lett.} \textbf{\bibinfo{volume}{94}},
  \bibinfo{pages}{167201} (\bibinfo{year}{2005});
\bibinfo{author}{\bibfnamefont{A.~A.} \bibnamefont{Kovalev}},
  \bibinfo{author}{\bibfnamefont{G.~E.~W.} \bibnamefont{Bauer}},
  \bibnamefont{and} \bibinfo{author}{\bibfnamefont{A.}~\bibnamefont{Brataas}},
  \bibinfo{journal}{Phys. Rev. B} \textbf{\bibinfo{volume}{75}},
  \bibinfo{pages}{014430} (\bibinfo{year}{2007}).


\bibitem[{\citenamefont{Jaafar and Chudnovsky}(2009)}]{Jaafar:Jun2009}
\bibinfo{author}{\bibfnamefont{R.}~\bibnamefont{Jaafar}} \bibnamefont{and}
  \bibinfo{author}{\bibfnamefont{E.~M.} \bibnamefont{Chudnovsky}},
  \bibinfo{journal}{Phys. Rev. Lett.} \textbf{\bibinfo{volume}{102}},
  \bibinfo{pages}{227202} (\bibinfo{year}{2009});
\bibinfo{author}{\bibfnamefont{R.}~\bibnamefont{Jaafar}},
  \bibinfo{author}{\bibfnamefont{E.~M.} \bibnamefont{Chudnovsky}},
  \bibnamefont{and} \bibinfo{author}{\bibfnamefont{D.~A.}
  \bibnamefont{Garanin}}, \bibinfo{journal}{Europhys. Lett.}
  \textbf{\bibinfo{volume}{89}}, \bibinfo{pages}{27001} (\bibinfo{year}{2010}).


\bibitem[{\citenamefont{Jaynes and Cummings}(1963)}]{Jaynes:jan.1963}
\bibinfo{author}{\bibfnamefont{E.}~\bibnamefont{Jaynes}} \bibnamefont{and}
  \bibinfo{author}{\bibfnamefont{F.}~\bibnamefont{Cummings}},
  \bibinfo{journal}{Proc. IEEE} \textbf{\bibinfo{volume}{51}},
  \bibinfo{pages}{89 } (\bibinfo{year}{1963});
\bibinfo{author}{\bibfnamefont{E.~K.} \bibnamefont{Irish}},
  \bibinfo{author}{\bibfnamefont{J.}~\bibnamefont{Gea-Banacloche}},
  \bibinfo{author}{\bibfnamefont{I.}~\bibnamefont{Martin}}, \bibnamefont{and}
  \bibinfo{author}{\bibfnamefont{K.~C.} \bibnamefont{Schwab}},
  \bibinfo{journal}{Phys. Rev. B} \textbf{\bibinfo{volume}{72}},
  \bibinfo{pages}{195410} (\bibinfo{year}{2005}).


\bibitem[{Not({\natexlab{a}})}]{Note}
\bibinfo{note}{For the ground state, there is only one
  maximum smeared by quantum fluctuations in the torsion angle, which makes it
  difficult to observe complete suppression of tunneling by interference; in this case, tunneling is gradually suppressed as quantum
  fluctuations become stronger.}

\bibitem[{\citenamefont{Wernsdorfer et~al.}(2000)\citenamefont{Wernsdorfer,
  Sessoli, Caneschi, Gatteschi, Cornia, and Mailly}}]{Wernsdorfer:MAY2000}
\bibinfo{author}{\bibfnamefont{W.}~\bibnamefont{Wernsdorfer}},
  \bibinfo{author}{\bibfnamefont{R.}~\bibnamefont{Sessoli}},
  \bibinfo{author}{\bibfnamefont{A.}~\bibnamefont{Caneschi}},
  \bibinfo{author}{\bibfnamefont{D.}~\bibnamefont{Gatteschi}},
  \bibinfo{author}{\bibfnamefont{A.}~\bibnamefont{Cornia}}, \bibnamefont{and}
  \bibinfo{author}{\bibfnamefont{D.}~\bibnamefont{Mailly}},
  \bibinfo{journal}{J. Appl. Phys.} \textbf{\bibinfo{volume}{87}},
  \bibinfo{pages}{5481} (\bibinfo{year}{2000}).

\bibitem[{\citenamefont{Perelomov}(1986)}]{Perelomov:1986}
\bibinfo{author}{\bibfnamefont{A.}~\bibnamefont{Perelomov}},
  \emph{\bibinfo{title}{Generalized Coherent States and Their Applications}}
  (\bibinfo{publisher}{Springer, New York}, \bibinfo{year}{1986}).

\bibitem[{\citenamefont{Chudnovsky and Gunther}(1988)}]{Chudnovsky:Feb1988}
\bibinfo{author}{\bibfnamefont{E.~M.} \bibnamefont{Chudnovsky}}
  \bibnamefont{and} \bibinfo{author}{\bibfnamefont{L.}~\bibnamefont{Gunther}},
  \bibinfo{journal}{Phys. Rev. Lett.} \textbf{\bibinfo{volume}{60}},
  \bibinfo{pages}{661} (\bibinfo{year}{1988}).

\bibitem[{\citenamefont{Loss et~al.}(1992)\citenamefont{Loss, DiVincenzo, and
  Grinstein}}]{Loss:Nov1992}
\bibinfo{author}{\bibfnamefont{D.}~\bibnamefont{Loss}},
  \bibinfo{author}{\bibfnamefont{D.~P.} \bibnamefont{DiVincenzo}},
  \bibnamefont{and}
  \bibinfo{author}{\bibfnamefont{G.}~\bibnamefont{Grinstein}},
  \bibinfo{journal}{Phys. Rev. Lett.} \textbf{\bibinfo{volume}{69}},
  \bibinfo{pages}{3232} (\bibinfo{year}{1992});
\bibinfo{author}{\bibfnamefont{M.~N.} \bibnamefont{Leuenberger}}
  \bibnamefont{and} \bibinfo{author}{\bibfnamefont{D.}~\bibnamefont{Loss}},
  \bibinfo{journal}{Phys. Rev. B} \textbf{\bibinfo{volume}{63}},
  \bibinfo{pages}{054414} (\bibinfo{year}{2001}).

\bibitem[{Not({\natexlab{b}})}]{Note1}
\bibinfo{note}{By eliminating $z_i$ these equations correspond to Lagrangian
  $\mathcal L = I \dot \varphi^2/2-K\varphi^2/2+S\hbar\dot \phi \cos
  \theta+S\hbar \dot \varphi \sin \theta \cos \varphi-E$ in which
  $\varphi=\alpha z_r/S$. This Lagrangian has been derived in M. F. O'Keeffe
  and E. M. Chudnovsky, arXiv:1011.3134v1 without the term $-K\varphi^2/2$.}

\bibitem[{\citenamefont{Cahill and Glauber}(1969)}]{Cahill:1969}
\bibinfo{author}{\bibfnamefont{K.~E.} \bibnamefont{Cahill}} \bibnamefont{and}
  \bibinfo{author}{\bibfnamefont{R.~J.} \bibnamefont{Glauber}},
  \bibinfo{journal}{Phys. Rev.} \textbf{\bibinfo{volume}{177}},
  \bibinfo{pages}{1857} (\bibinfo{year}{1969}).

\bibitem[{\citenamefont{Chudnovsky and Garanin}(2010)}]{Chudnovsky:Jun2010}
\bibinfo{author}{\bibfnamefont{E.~M.} \bibnamefont{Chudnovsky}}
  \bibnamefont{and} \bibinfo{author}{\bibfnamefont{D.~A.}
  \bibnamefont{Garanin}}, \bibinfo{journal}{Phys. Rev. B}
  \textbf{\bibinfo{volume}{81}}, \bibinfo{pages}{214423}
  (\bibinfo{year}{2010}).

\bibitem[{\citenamefont{Rocheleau et~al.}(2010)\citenamefont{Rocheleau, Ndukum,
  Macklin, Hertzberg, Clerk, and Schwab}}]{Rocheleau:jan2010}
\bibinfo{author}{\bibfnamefont{T.}~\bibnamefont{Rocheleau}},
  \bibinfo{author}{\bibfnamefont{T.}~\bibnamefont{Ndukum}},
  \bibinfo{author}{\bibfnamefont{C.}~\bibnamefont{Macklin}},
  \bibinfo{author}{\bibfnamefont{J.~B.} \bibnamefont{Hertzberg}},
  \bibinfo{author}{\bibfnamefont{A.~A.} \bibnamefont{Clerk}}, \bibnamefont{and}
  \bibinfo{author}{\bibfnamefont{K.~C.} \bibnamefont{Schwab}},
  \bibinfo{journal}{Nature} \textbf{\bibinfo{volume}{463}}, \bibinfo{pages}{72}
  (\bibinfo{year}{2010}).

\bibitem[{\citenamefont{Meyer et~al.}(2005)\citenamefont{Meyer, Paillet, and
  Roth}}]{Meyer:SEP2005}
\bibinfo{author}{\bibfnamefont{J.~C.} \bibnamefont{Meyer}},
  \bibinfo{author}{\bibfnamefont{M.}~\bibnamefont{Paillet}}, \bibnamefont{and}
  \bibinfo{author}{\bibfnamefont{S.}~\bibnamefont{Roth}},
  \bibinfo{journal}{Science} \textbf{\bibinfo{volume}{309}},
  \bibinfo{pages}{1539} (\bibinfo{year}{2005}).

\bibitem[{\citenamefont{Moushi et~al.}(2009)\citenamefont{Moushi, Stamatatos,
  Wernsdorfer, Nastopoulos, Christou, and Tasiopoulos}}]{Moushi:SEP2009}
\bibinfo{author}{\bibfnamefont{E.~E.} \bibnamefont{Moushi}},
  \bibinfo{author}{\bibfnamefont{T.~C.} \bibnamefont{Stamatatos}},
  \bibinfo{author}{\bibfnamefont{W.}~\bibnamefont{Wernsdorfer}},
  \bibinfo{author}{\bibfnamefont{V.}~\bibnamefont{Nastopoulos}},
  \bibinfo{author}{\bibfnamefont{G.}~\bibnamefont{Christou}}, \bibnamefont{and}
  \bibinfo{author}{\bibfnamefont{A.~J.} \bibnamefont{Tasiopoulos}},
  \bibinfo{journal}{Inorg. Chem.} \textbf{\bibinfo{volume}{48}},
  \bibinfo{pages}{5049} (\bibinfo{year}{2009}).

\bibitem[{\citenamefont{Barraza-Lopez et~al.}(2009)\citenamefont{Barraza-Lopez,
  Park, Garc\'\i{}a-Su\'arez, and Ferrer}}]{Barraza-Lopez:Jun2009}
\bibinfo{author}{\bibfnamefont{S.}~\bibnamefont{Barraza-Lopez}},
  \bibinfo{author}{\bibfnamefont{K.}~\bibnamefont{Park}},
  \bibinfo{author}{\bibfnamefont{V.}~\bibnamefont{Garc\'\i{}a-Su\'arez}},
  \bibnamefont{and} \bibinfo{author}{\bibfnamefont{J.}~\bibnamefont{Ferrer}},
  \bibinfo{journal}{Phys. Rev. Lett.} \textbf{\bibinfo{volume}{102}},
  \bibinfo{pages}{246801} (\bibinfo{year}{2009}).

\bibitem[{\citenamefont{del Barco et~al.}(1999)\citenamefont{del Barco,
  Vernier, Hernandez, Tejada, Chudnovsky, Molins, and
  Bellessa}}]{Barco:SEP1999}
\bibinfo{author}{\bibfnamefont{E.}~\bibnamefont{del Barco}},
  \bibinfo{author}{\bibfnamefont{N.}~\bibnamefont{Vernier}},
  \bibinfo{author}{\bibfnamefont{J.~M.} \bibnamefont{Hernandez}},
  \bibinfo{author}{\bibfnamefont{J.}~\bibnamefont{Tejada}},
  \bibinfo{author}{\bibfnamefont{E.~M.} \bibnamefont{Chudnovsky}},
  \bibinfo{author}{\bibfnamefont{E.}~\bibnamefont{Molins}}, \bibnamefont{and}
  \bibinfo{author}{\bibfnamefont{G.}~\bibnamefont{Bellessa}},
  \bibinfo{journal}{Europhys. Lett.} \textbf{\bibinfo{volume}{47}},
  \bibinfo{pages}{722} (\bibinfo{year}{1999}).

\end{thebibliography}

\end{document}